\newtheorem{theorem}{Theorem}
\newtheorem{proposition}{Proposition}

\newtheorem{p-assumption}{Preference Assumption}
\newtheorem{gr-assumption}{Satisfaction Assumption}
\newtheorem{e-assumption}{Monotonicity Assumption}
\newtheorem{e-assumption-p}{Parametric Assumption}
\newtheorem{cr-assumption}{Coherent-risk Assumption}
\newtheorem{definition}{Definition}

\newtheorem{conjecture}{Conjecture}

\documentclass[12pt]{article}

\usepackage{amsmath}
\usepackage{amssymb}
\usepackage{amsfonts}
\usepackage{mathrsfs}

\usepackage{graphicx}
\newcommand{\qed}{{\mbox{} \hspace*{\fill}{\vrule height5pt width4pt
			depth0pt}}\\}

\def\M{\hspace*{0.75em}}

\begin{document}

\title{A Conjecture Related to the Traveling Salesman Problem}
\author{Jian Yang\\
Department of Management Science and Information Systems\\
Business School, Rutgers University\\
Newark, NJ 07102\\
Email: jyang@business.rutgers.edu}
\date{September 2000; revised, May 2002, October 2021}
\maketitle

\begin{abstract}

We show that certain ways of solving some combinatorial optimization problems can be understood as using query planes to divide the space of problem instances into polyhedra that could fit into those that characterize the problem's various solutions. This viewpoint naturally leads to a splinter-proneness property that is then shown to be responsible for the hardness of the concerned problem. We conjecture that the $NP$-equivalent traveling salesman problem (TSP) has this property and hence is hard to solve to a certain extent. 

\vspace{3mm}
\end{abstract}

\newpage 

\section{Introduction}


Traditionally, people view a combinatorial optimization problem (COP) as a mapping from a problem instance to one of its optimal solutions. For example, the traveling salesman problem (TSP) is expressed as one seeking the least costly tour when given information about the underlying cities; see \cite{tsp}. An alternative view might help us to better understand why one single algorithm for a particular COP is able to respond to the possibly infinite varieties of instances for that COP and always reach the correct answer. 

Rather than an instance-wide one, a COP's feasible solution can be treated as a problem-wide concept. Also, the relationship of an
instance to its optimal solution is less of ``corresponds to'' than of
``belongs to''. For TSP with $n+1$ cities, for instance, there are $n!$ feasible solutions, namely, the $n!$ sequences of visiting the cities. An instance merely expresses the information on pairwise distances of the cities and it belongs to {\em the set of instances that share the same optimal solution with it}. For a given feasible solution $s$, we shall later call the set containing every instance that has $s$ as an optimal solution the $S$-set of $s$.

With the above understanding of a COP, an algorithm for it can then be viewed as a history-dependent sequence of divisions in the instance space. It first divides the instance space into two halves, say the +1 half and the -1 half. Then, it divides the +1 half into the +1+1 half and the +1-1 half; also, it divides the -1 half into the -1+1 half and the -1-1 half. The algorithm would be considered as having solved the problem when the process goes on both accurately and long enough so that {\em every set resulting from its history of divisions is entirely inside one $S$-set}. In other words, an algorithm for a COP is just a binary tree with every node being a set of instances, with every branching being a division of the set that is the father node into two sets that are the two children nodes, with its root being the entire instance space, and with each of its leaves being entirely inside one of the $S$-sets. Hence, given an instance, the algorithm executes the sequence of divisions that
define the leaf to which the instance belongs and then reports the
feasible solution whose $S$-set contains this leaf as the optimal
solution for this instance. 

For a variety of COPs, which we will later call RILCOPs, we further
realize that each of their $S$-sets, when being mildly extrapolated if necessary, is merely a polyhedron. For these COPs, observation and intuition tempt us to surmise that as far as efficient solvings are concerned, only needed are SIMPLE algorithms that divide the instance space by hyperplanes that serve as queries. So that the problem's $S$-sets eventually contain all polyhedra made up of half spaces resulting from query planes, all areas of their faces need to be scraped by some query planes. The difficulty of solving an RILCOP is thus linked to the hardness of finding query planes that position suitably with its $S$-sets. Especially, we stumble upon a {\em splinter-proneness} property and show that an RILCOP possessing it would not be solvable by any SIMPLE algorithm in polynomial time. We also conjecture that TSP is splinter-prone. Since the problem is not only $NP$-hard, but also $NP$-easy \cite{Garey and Johnson}, this conjecture serves as a sufficient condition for the well-known and puzzling conjecture $P\neq NP$ if a SIMPLE algorithm is all that is  needed to solve TSP.


The rest of the paper is organized as follows. We define RILCOPs and SIMPLE algorithms in Section~\ref{sec2} and elaborate on how a former can be solved by a latter in Section~\ref{sec3}. In Section~\ref{sec4}, we propose the splinter-proneness concept and demonstrate its sufficient-condition status for hardness. In Section~\ref{sec5}, we make our conjecture for TSP and discuss relevant topics. The paper is concluded in Section~\ref{sec6}.

\section{RILCOPs and SIMPLE Algorithms}\label{sec2}

Consider a COP with the following features.\\
(i) Given a certain size parameter $n$, its instance space can be represented by a fixed-dimensional space of rational numbers and there is a finite set of feasible solutions;\\
(ii) Under a feasible solution, there is a linear objective value function of the instance; also, the COP seeks to minimize the objective. \\
The assignment problem (AP) and TSP both belong to this type of rational-input linear COPs (RILCOPs). We have insisted on (ii)'s single objective version for ease of presentation. A slight generalization without altering the essence would allow RILCOP to cover the shortest path problem (SP) and the minimal spanning tree problem (MST).

For instance, TSP of size $3$ can have each of its instances represented
by a $3\cdot (3+1)=12$-component rational vector: 
\begin{equation}\label{tsp-3-c}
{\bf c}\equiv(c_{01},c_{02},c_{03},c_{10},c_{12},c_{13},c_{20},c_{21},c_{23},
c_{30},c_{31},c_{32})^{T}.
\end{equation}
Common to all instances are $3!=6$ feasible solutions: 
\begin{equation}\label{tsp-3-s}\left\{\begin{array}{l}
(3,0)=\mbox{ tour }0\longrightarrow 1\longrightarrow 2\longrightarrow 3\longrightarrow 0,\\
(3,1)=\mbox{ tour }0\longrightarrow 1\longrightarrow 3\longrightarrow 2\longrightarrow 0,\\                                                        
(3,2)=\mbox{ tour }0\longrightarrow 2\longrightarrow 1\longrightarrow 3
\longrightarrow 0,\\                                                        
(3,3)=\mbox{ tour }0\longrightarrow 2\longrightarrow 3\longrightarrow 1
\longrightarrow 0,\\                                                        
(3,4)=\mbox{ tour }0\longrightarrow 3\longrightarrow 1\longrightarrow 2
\longrightarrow 0,\\                                                        
(3,5)=\mbox{ tour }0\longrightarrow 3\longrightarrow 2\longrightarrow 1
\longrightarrow 0.
\end{array}\right.
\end{equation}
To each feasible solution $(3,s)$, there would correspond a linear objective function $({\bf x}_{\mbox{tsp}}(3,s))^T\cdot{\bf c}$. For example, ${\bf x}_{\mbox{tsp}}(3,0)\equiv (1,0,0,0,1,0,0,0,1,1,0,0)^T$ and hence
\begin{equation}\label{tsp-3-0}
({\bf x}_{\mbox{tsp}}(3,0))^T\cdot {\bf c}=c_{01}+c_{12}+c_{23}+c_{30}.
\end{equation}

For an RILCOP ${\cal P}$ of a certain size $n\in \mathbb N$ where $\mathbb N\equiv \{1,2,...\}$ stands for the space of natural numbers, we denote its entire instance space by $I_{\cal P}(n)$. The latter is just $\mathbb Q^{D_{\cal P}(n)}$ itself, where $\mathbb Q$ is the space of rationals and $D_{\cal P}(n)$ is a polynomial of $n$, or a polyhedron inside the space. For instance, $I_{\mbox{tsp}}(n)=\mathbb Q^{D_{\mbox{tsp}}(n)}$ with $D_{\mbox{tsp}}(n)=n\cdot (n+1)$. We also denote the problem's number of distinct feasible solutions by $S_{\cal P}(n)$ and just name them $(n,s)$ with $s=0,1,...,S_{\cal P}(n)-1$. The $\mathbb Q^{D_{\cal P}(n)}$-vector that gives an $(n,s)$-corresponding linear objective would be denoted by ${\bf x}_{\cal P}(n,s)$. Given a feasible solution $(n,s)$, we denote the set of instances that have $s$ as one of its optimal solutions by $I^{S}_{\cal P}(n,s)$, which amounts to 
\begin{equation}\label{def-solution-set}
\left\{{\bf c}\in I_{\cal P}(n):\;\left({\bf x}_{\cal P}(n,s)-{\bf x}_{\cal P}(n,s')\right)^T\cdot{\bf c}\leq 0\hspace*{.2in}\forall s'=0,...,s-1,s+1,...,S_{\cal P}(n)-1\right\}.
\end{equation}
By~(\ref{tsp-3-s}) to~(\ref{def-solution-set}), the $S$-set $I^S_{\mbox{tsp}}(3,0)$ would contain all ${\bf c}\in \mathbb Q^{12}$ of~(\ref{tsp-3-c}) that satisfy
\[ \left\{\begin{array}{l}
\left({\bf x}_{\mbox{tsp}}(3,0)-{\bf x}_{\mbox{tsp}}(3,1)\right)^T\cdot {\bf c}=(c_{01}+c_{12}+c_{23}+c_{30})-(c_{01}+c_{13}+c_{32}+c_{20})\\
\;\;\;\;\;\;\;\;\;\;\;\;\;\;\;\;\;\;\;\;\;\;\;\;\;\;\;\;\;\;\;\;\;\;\;\;\;\;\;\;\;\;\;\;\;\;=c_{12}-c_{13}-c_{20}+c_{23}+c_{30}-c_{32}\leq 0,\\
\left({\bf x}_{\mbox{tsp}}(3,0)-{\bf x}_{\mbox{tsp}}(3,2)\right)^T\cdot {\bf c}=(c_{01}+c_{12}+c_{23}+c_{30})-(c_{02}+c_{21}+c_{13}+c_{30})\\
\;\;\;\;\;\;\;\;\;\;\;\;\;\;\;\;\;\;\;\;\;\;\;\;\;\;\;\;\;\;\;\;\;\;\;\;\;\;\;\;\;\;\;\;\;\;=c_{01}-c_{02}+c_{12}-c_{13}-c_{21}+c_{23}\leq 0,\\
\left({\bf x}_{\mbox{tsp}}(3,0)-{\bf x}_{\mbox{tsp}}(3,3)\right)^T\cdot {\bf c}=(c_{01}+c_{12}+c_{23}+c_{30})-(c_{02}+c_{23}+c_{31}+c_{10})\\
\;\;\;\;\;\;\;\;\;\;\;\;\;\;\;\;\;\;\;\;\;\;\;\;\;\;\;\;\;\;\;\;\;\;\;\;\;\;\;\;\;\;\;\;\;\;=c_{01}-c_{02}-c_{10}+c_{12}+c_{30}-c_{31}\leq 0,\\
\left({\bf x}_{\mbox{tsp}}(3,0)-{\bf x}_{\mbox{tsp}}(3,4)\right)^T\cdot {\bf c}=(c_{01}+c_{12}+c_{23}+c_{30})-(c_{03}+c_{31}+c_{12}+c_{20})\\
\;\;\;\;\;\;\;\;\;\;\;\;\;\;\;\;\;\;\;\;\;\;\;\;\;\;\;\;\;\;\;\;\;\;\;\;\;\;\;\;\;\;\;\;\;\;=c_{01}-c_{03}-c_{20}+c_{23}+c_{30}-c_{31}\leq 0,\\
\left({\bf x}_{\mbox{tsp}}(3,0)-{\bf x}_{\mbox{tsp}}(3,5)\right)^T\cdot {\bf c}=(c_{01}+c_{12}+c_{23}+c_{30})-(c_{03}+c_{32}+c_{21}+c_{10})\\
\;\;\;\;\;\;\;\;\;\;\;\;\;\;\;\;\;\;\;\;\;\;\;\;\;\;\;\;\;\;\;\;\;\;\;\;\;\;\;\;\;\;\;\;\;\;=c_{01}-c_{03}-c_{10}+c_{12}-c_{21}+c_{23}+c_{30}-c_{32}\leq 0.
\end{array}\right.\]
                                                        
For TSP, as for many other RILCOPs, there does not seem to be any need to go through operations other than MULTIPLICATION of a component by a constant rational, ADDITION between two rationals resulting from the first operation, COMPARISON of any resulting rational with zero, and OTHER operations that do not involve the instance. 
In the Hungarian method for solving AP, OTHER operations include solving an entire unweighted maximal matching problem. At a certain stage of the Hungarian method, the 0-1 information indicating which unweighted maximal matching problem is to be solved is indeed instance-dependent; however, the solving itself does not involve any of the first three operations. We consider an algorithm SIMPLE when it resorts to only the above four operations. 

In essence, a SIMPLE algorithm is one that never breaks an instance by non-arithmetic operations and always performs linear operations on it. Dijkstra's algorithms for SP and MST \cite{Dijkstra}, 
Kruskal's algorithm for the latter problem \cite{Kruskal}, the Hungarian method for AP \cite{Kuhn} are all examples of SIMPLE algorithms.

There is a somewhat ironic counter example. Linear programming (LP) can be viewed as a COP, but not an RILCOP. In addition, the simplex method is not a SIMPLE algorithm even when COMPARISON is allowed to be against any nonzero rational. A standard-form LP with $m$ rows and $n$ columns ($n\geq
m$) can be represented by a rational vector with $n+mn+m$ components, say 
\[ {\bf e}\equiv(c_{1},...,c_{n},a_{11},...,a_{1n},...,a_{m1},...,a_{mn},b_{1},...,b_{m})^T,\]
where $c_{j}$'s are the conventional objective coefficients, $a_{ij}$'s the constraint coefficients, and $b_{i}$'s the right-hand side coefficients. Let $(n,m,0)$, $(n,m,1)$, $...$, $(n,m,n!/(m!\cdot (n-m)!)-1)$ stand for all the $n!/(m!\cdot (n-m)!)$ bases without regard to their feasibility. The closest we can model LP as an RILCOP is to write its objective at each $(n,m,s)$-feasible solution as 
\[ ({\bf x}_{\mbox{lp}}(n,m,s,{\bf e}))^T\cdot {\bf e}=\sum_{j=1}^n x_{\mbox{lp},j}(n,m,s,{\bf e})\cdot c_j\]
where ${\bf x}_{\mbox{lp}}(n,m,s,{\bf e})\equiv (x_{\mbox{lp},1}(n,m,s,{\bf e}),...,x_{\mbox{lp},n}(n,m,s,{\bf e}))^T$ is a basic feasible solution for the basis $(n,m,s)$ under the 
instance ${\bf e}$ when this is possible and $+\infty$ when $(n,m,s)$ is not a feasible basis for ${\bf e}$. Due to the clear ${\bf e}$-dependence of ${\bf x}_{\mbox{lp}}(n,m,s,{\bf e})$, LP would not actually be an RILCOP. The simplex method is not SIMPLE either. It has the occasion to perform divisions with both operands being components of the instance. For RILCOPs, we have not seen any attempt to solve them using non-SIMPLE algorithms. 

\section{Solving an RILCOP ${\cal P}$ by a SIMPLE ${\cal A}$}\label{sec3}

When running a SIMPLE algorithm ${\cal A}$ on an RILCOP ${\cal P}$ of size $n$, all instances in $I_{\cal P}(n)$ would experience the same sequence of operations up to the first COMPARISON in ${\cal A}$. It would be about whether $({\bf q}_{\cal A}(n))^{T}\cdot {\bf c}\leq 0$ for some ${\bf q}_{\cal A}(n)\in \mathbb Q^{D_{\cal P}(n)}$. According to the answer being {\em yes} or {\em no}, this first COMPARISON would divide $I_{\cal P}(n)$ into two polyhedra $I_{{\cal P},{\cal A}}(n,+1)$ and $I_{{\cal P},{\cal A}}(n,-1)$. All instances in $I_{{\cal P},{\cal A}}(n,+1)$ would in turn experience the same sequence of operations up to the next COMPARISON. It would be about whether $({\bf q}_{\cal A}(n,+1))^T\cdot {\bf c}\leq 0$ for some ${\bf q}_{\cal A}(n,+1)\in \mathbb Q^{D_{\cal P}(n)}$. Again, according to the answer being {\em yes} or {\em no}, this COMPARISON would divide $I_{{\cal P},{\cal A}}(n,+1)$ into $I_{{\cal P},{\cal A}}(n,+1,+1)$ and $I_{{\cal P},{\cal A}}(n,+1,-1)$. Similar statements can be made about $I_{{\cal P},{\cal A}}(n,-1)$ and until the RILCOP has been solved, for any $I_{{\cal P},{\cal A}}(n,{\bf b})$ where ${\bf b}$ is a binary sequence of $+1$'s and $-1$'s. When $I_{{\cal P},{\cal A}}(n,{\bf b})\subseteq I^S_{\cal P}(n,s)$ for some binary sequence ${\bf b}$ and $s=0,1,...,S_{\cal P}(n)-1$, the answers $b_l$ to queries of whether $({\bf q}_{\cal A}(n,b_1,...,b_{l-1}))^T\cdot{\bf c}\leq 0$ would be sufficient to detect that ${\bf c}\in I^S_{\cal P}(n,s)$. 

Without loss of generality, suppose $I_{\cal P}(n)$ is just $\mathbb Q^{D_{\cal P}(n)}$ itself. Then, the entire RILCOP ${\cal P}$ can be represented by the sequences 
\begin{equation}\label{def-p-problem}
(D_{\cal P}(n))_{n\in\mathbb N},\;\;\;(S_{\cal P}(n))_{n\in\mathbb N},\;\;\; \mbox{ and }\;\;\left(\left({\bf x}_{\cal P}(n,s)\right)_{s=0,1,...,S_{\cal P}(n)-1}\right)_{n\in\mathbb N},
\end{equation}
where each ${\bf x}_{\cal P}(n,s)\in \mathbb Q^{D_{\cal P}(n)}$. Also, all aforementioned polyhedra can be represented by the ensembles of inequalities of the form ${\bf f}^T\cdot {\bf c}\leq 0$ that help define them, where the ${\bf f}$'s are vectors in $\mathbb Q^{D_{\cal P}(n)}$. Now each $I_{\cal P}(n)$ is just the null ensemble $()$ and according to~(\ref{def-solution-set}), each $I^S_{\cal P}(n,s)$ is just the $(S_{\cal P}(n)-1)$-member $D_{\cal P}(n)$-dimensional ensemble
\begin{equation}\label{outdated}
\left({\bf x}_{\cal P}(n,s)-{\bf x}_{\cal P}(n,s')\right)_{s'=0,...,s-1,s+1,...,S_{\cal P}(n)-1}.
\end{equation}

To describe an algorithm ${\cal A}$ more precisely, let $\mathscr B_l$ be the set of all $2^l$ number of $l$-long $(+1,-1)$-valued binary strings. We consider a collection ${\cal B}$ of strings {\em tree-like} when for any ${\bf b}\equiv (b_1,...,b_l)\in {\cal B}$ with $l=1,2,...$, both the ``parent'' string ${\bf b}\ominus (b_l)\equiv (b_1,...,b_{l-1})$ and ``sibling'' string ${\bf b}\ominus (b_l)\oplus (-b_l)\equiv (b_1,...,b_{l-1},-b_l)$ would be in ${\cal B}$ as well. We consider a SIMPLE algorithm ${\cal A}$ {\em compatible} with an RILCOP ${\cal P}$ defined through~(\ref{def-p-problem}) when for any $n\in \mathbb N$, there is a tree-like collection ${\cal B}_{\cal A}(n)\subset \bigcup_{l=0}^{+\infty}\mathscr B_l$, so that for any string ${\bf b}\in {\cal B}_{\cal A}(n)$, there exists a query ${\bf q}_{\cal A}(n,{\bf b})\in \mathbb Q^{D_{\cal P}(n)}$. That is, a ${\cal P}$-compatible  ${\cal A}$ is basically  
\[ ({\cal B}_{\cal A}(n))_{n\in\mathbb N}\;\;\;\mbox{ and }\;\;\left(\left({\bf q}_{\cal A}(n,{\bf b})\right)_{{\bf b}\in {\cal B}_{\cal A}(n)}\right)_{n\in\mathbb N},\]
where each ${\cal B}_{\cal A}(n)$ is tree-like and each ${\bf q}_{\cal A}(n,{\bf b})\in\mathbb Q^{D_{\cal P}(n)}$. 

Given an $l$-string ${\bf b}\equiv (b_1,...,b_l)$, let us denote by ${\bf b}_{[l',l'']}$ the sub-string $(b_m)_{m=l',...,l''}$. Note in algorithm ${\cal A}$ at size parameter $n$, queries associated with a given string ${\bf b}\equiv (b_1,...,b_l)\in {\cal B}_{\cal A}(n)$ are ${\bf q}_{\cal  A}(n)$, ${\bf q}_{\cal A}(n,b_1)$, $...$, ${\bf q}_{\cal A}(n,{\bf b})$ and the ultimate two polyhedra resulting from these queries are $(b_j\cdot{\bf q}_{\cal A}(n,{\bf b}_{[1,j-1]}))_{j=1,2,...,l}\oplus(+{\bf q}_{\cal A}(n,{\bf b}))$ and $(b_j\cdot{\bf q}_{\cal A}(n,{\bf b}_{[1,j-1]}))_{j=1,2,...,l}\oplus(-{\bf q}_{\cal A}(n,{\bf b}))$, where $\oplus$ has been used for concatenation. Let us reiterate our convention that the former would stand for the polyhedron within $\mathbb Q^{D_{\cal P}(n)}$ where all vectors ${\bf c}$ satisfy 
\[ \left(b_1\cdot {\bf q}_{\cal A}(n)\right)^T\cdot {\bf c}\leq 0,\]
which actually means $({\bf q}_{\cal A}(n))^T\cdot {\bf c}\leq 0$ when $b_1=+1$ and $({\bf q}_{\cal A}(n))^T\cdot {\bf c}\geq 0$ when $b_1=-1$,  
\[ \left(b_2\cdot {\bf q}_{\cal A}(n,b_1)\right)^T\cdot {\bf c}\leq 0,\;\;\;\;\;\;......,\;\;\;\;\;\;
\left(b_l\cdot {\bf q}_{\cal A}(n,{\bf b}_{[1,l-1]})\right)^T\cdot {\bf c}\leq 0,\]
and ultimately, $({\bf q}_{\cal A}(n,{\bf b}))^T\cdot {\bf c}\leq 0$. The latter polyhedron is different only in its opposite sign for the last linear inequality. 

Given a tree-like binary string collection ${\cal B}$, we can define its leaf sub-collection ${\cal B}^0$ as one that contains all ${\bf b}^0\in {\cal B}$ whose children ${\bf b}^0\oplus(+1)$ and ${\bf b}^0\oplus(-1)$ are no longer in ${\cal B}$. Indeed, the latter could be understood as being generated by ${\cal B}^0$ as every ${\bf b}\in {\cal B}$ is obtainable by conducting a number of $\ominus$ operations starting from some ${\bf b}^0\in {\cal B}^0$. From the above, we may consider a ${\cal P}$-compatible algorithm ${\cal A}$ as being capable of {\em solving} ${\cal P}$ when for any $n\in\mathbb N$ and any ${\bf b}\equiv (b_1,...,b_l)$ in the leaf sub-collection $({\cal B}_{\cal A}(n))^0$ of the tree-like ${\cal B}_{\cal A}(n)$, there would exist some $s_+,s_-=0,1,...,S_{\cal P}(n)-1$ such that 
\[ \left(b_j\cdot {\bf q}_{\cal A}(n,{\bf b}_{[1,j-1]})\right)_{j=1,2,...,l}\oplus (\pm{\bf q}_{\cal A}(n,{\bf b }))\subseteq \left({\bf x}_{\cal P}(n,s_{\pm})-{\bf x}_{\cal P}(n,s')\right)_{s'=0,...,s_{\pm}-1,s_{\pm}+1,...,S_{\cal P}(n)-1}.\]

By Farkas' Lemma, the above would be translatable into the existence of $2l\cdot (S_{\cal P}(n)-1)$ positive rationals $\lambda^{\pm}_{s'1},...,\lambda^{\pm}_{s',l-1},\lambda^{\pm}_{s'l}$ for $s'=0,...,s_{\pm}-1,s_{\pm}+1,...,S_{\cal P}(n)-1$ so that for either + or - and each such $s'\neq s_{\pm}$, we have the $\mathbb Q^{D_{\cal P}(n)}$-equation that
\begin{equation}\label{baffled}
{\bf x}_{\cal P}(n,s_{\pm})-{\bf x}_{\cal P}(n,s')=\sum_{j=1}^{l-1}\lambda^{\pm}_{s'j}b_j\cdot{\bf q}_{\cal A}(n,{\bf b}_{[1,j-1]})\pm \lambda^{\pm}_{s'l}\cdot {\bf q}_{\cal A}(n,{\bf b}).
\end{equation}

\section{Faces, Splinter-proneness, and a Conjecture}\label{sec4}

Among every of $I^S_{\cal P}(n,s)$'s boundaries that include vectors ${\bf c}$ that satisfy $\left({\bf x}_{\cal P}(n,s)-{\bf x}_{\cal P}(n,s')\right)^T\cdot{\bf c}=0$ for some fixed $s'=0,...,s-1,s+1,...,S_{\cal P}(n)-1$, the faces would stand out. Let $F_{\cal P}(n,s)\subseteq \{0,...,s-1,s+1,...,S_{\cal P}(n)-1\}$ be the set of $I^S_{\cal P}(n,s)$'s set of faces. Suppose $s'\in F_{\cal P}(n,s)$. Then for any positive rationals $\lambda_{s''}$ where $s''=0,1,...,S_{\cal P}(n)-1$ is neither $s$ nor $s'$, it would {\em not happen} that 
\[ {\bf x}_{\cal P}(n,s')=\sum_{s''=1\mbox{ to }S_{\cal P}(n)-1,\;s''\neq s,\;s''\neq s'}\lambda_{s''}\cdot {\bf x}_{\cal P}(n,s'').\]
Rather than~(\ref{outdated}), each $I_{\cal P}(n,s)$ can now be face-defined in the manner of
\[ \left({\bf x}_{\cal P}(n,s)-{\bf x}_{\cal P}(n,s')\right)_{s'\in F_{\cal P}(n,s)}.
\]


For the relationship between a query plane ${\bf q}\in \mathbb Q^d$ and a face-defined polyhedron $I\equiv ({\bf y}_f)_{f\in F}\subseteq \mathbb Q^d$ where $d$ is some finite dimension, $F$ is a finite set, and each ${\bf y}_f$ is a face of $I$, there exists a three-way categorization: \\
\indent\M (spL) ${\bf q}$ splinters $I$ into two halves, without even scraping any of its faces; more specifically, it would amount to the {\em non-existence of} an either all-positive or all-negative multiplier ensemble $(\lambda_f)_{f\in F}$ such that ${\bf q}=\sum_{f\in F}{\bf y}_f$;\\
\indent\M (sCp) ${\bf q}$ scrapes one of $I$'s faces; more specifically, it would mean the {\em existence of one single} $f\in F$ and some $\lambda_f\neq 0$ such that ${\bf q}=\lambda_f\cdot {\bf y}_f$;\\
\indent\M (Rsp) ${\bf q}$ remotely supports $I$ from one side without scraping; more specifically, it would mean the {\em existence of} an either all-positive or all-negative ensemble $(\lambda_f)_{f\in F}$ such that ${\bf q}=\sum_{f\in F} \lambda_f\cdot {\bf y}_f$ while without ${\bf q}$ being proportional to any one single ${\bf y}_f$ which also indicates that {\em at least two} $\lambda_f$'s in the multiplier ensemble are nonzero.  

\begin{figure}[h]
\centering
\includegraphics[width=1.0\columnwidth]{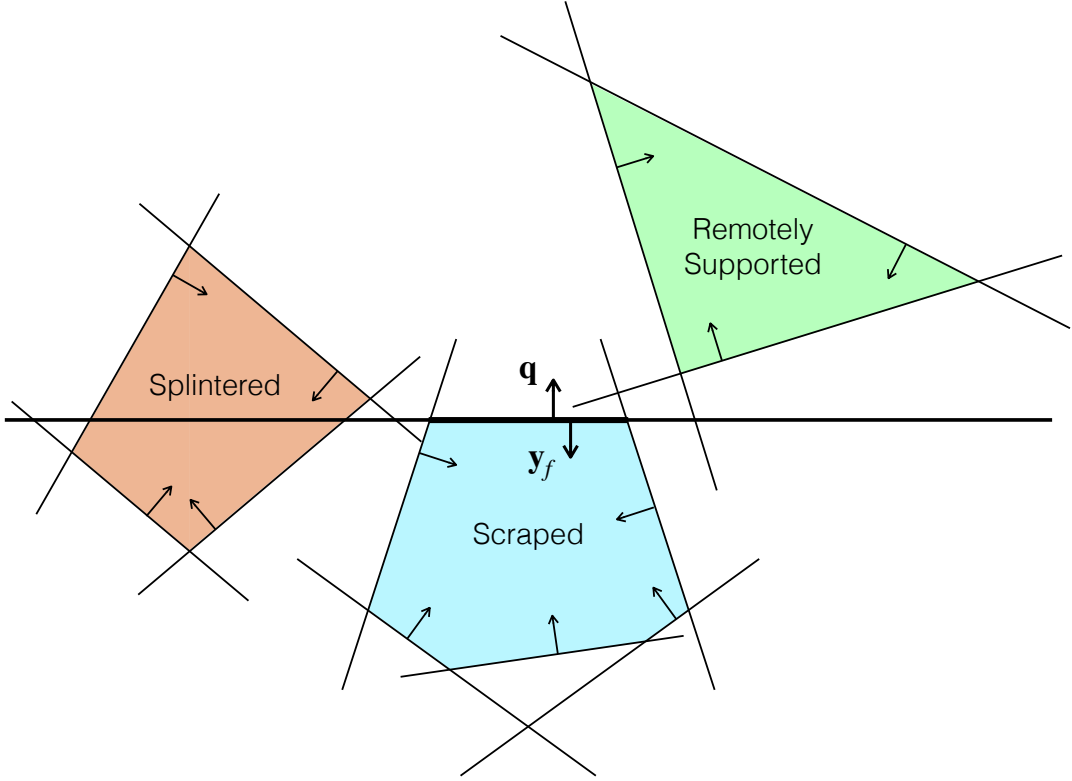}
\caption{An Illustration of the Three Categories}\label{fig1}
\end{figure}

These three categories are illustrated in Figure~\ref{fig1}.

The above concepts are also linked to a balancedness concept. We consider a query ${\bf q}\equiv (q_k)_{k=1,...,d}\in\mathbb Q^d$ balanced when $\sum_{k=1}^d q_k=0$. A polyhedron $I\equiv ({\bf y}_f)_{f\in F}$ would be deemed balanced when every of its defining queries ${\bf y}_f$ is balanced. 

\begin{proposition}\label{p-balanced}
An unbalanced query ${\bf q}$ would always splinter a balanced polyhedron $I\equiv (y_f)_{f\in F}$ into two halves.
\end{proposition}

\noindent{\bf Proof of Proposition~\ref{p-balanced}: }We can prove this by contradiction. Suppose an unbalanced query ${\bf q}$ did not splinter a balanced polyhedron $I\equiv (y_f)_{f\in F}$. Then from the definitions of (sCp) and (Rsp), there must exist an all-positive or all-negative ensemble $(\lambda_f)_{f\in F}$ such that ${\bf q}=\sum_{f\in F}\lambda_f\cdot{\bf y}_f$. If we sum over all components, we would then obtain 
\[ 0\neq \sum_{k=1}^d q_k=\sum_{k=1}^d\sum_{f\in F}\lambda_f\cdot y_{fk}=\sum_{f\in F}\lambda_f\cdot\sum_{k=1}^d y_{fk}=0,\]
which is apparently a contradiction. \qed 

Certainly, query-resulting polyhedra getting inside every $I^S_{\cal P}(n,s)$ in the fashion of~(\ref{baffled}) would be equivalent to {\em all faces having been scraped}. 
 
\begin{definition}\label{def-solved}
An RILCOP ${\cal P}$ would be considered as being {\em solved} by a compatible ${\cal A}$ when for any $n\in \mathbb N$, all face points of all $I^S_{\cal P}(n,s)$'s are to be scraped by query planes of ${\cal A}$. 
\end{definition} 

For ${\bf y}\neq {\bf 0}$, let us use the notation $\left({\bf y}|({\bf y}_j)_{j\in J}\right)$ to stand for the set of $\mathbb Q^d$-members ${\bf c}$ that satisfy ${\bf y}^T\cdot {\bf c}=0$ and ${\bf y}_j^{\;T}\cdot {\bf c}\leq 0$ for $j\in J$. For $\lambda\neq 0$, note $\left(\lambda\cdot {\bf y}|({\bf y}_j)_{i\in J}\right)$ would mean a polyhedron in $\mathbb Q^{d-1}$. Now more precisely, Definition~\ref{def-solved} means that, for any $n\in\mathbb N$, $s=0,1,...,S_{\cal P}(n)-1$, and $s'\in F_{\cal P}(n,s)$, the $(D_{\cal P}(n)-1)$-dimensional face
\begin{equation}\label{baba}
\left({\bf x}_{\cal P}(n,s)-{\bf x}_{\cal P}(n,s')|\left({\bf x}_{\cal P}(n,s)-{\bf x}_{\cal P}(n,s'')\right)_{s''\in F_{\cal P}(n,s)\setminus\{s'\}}\right)
\end{equation}  
would always be inside the union of potential faces 
\begin{equation}\label{mama}
\bigcup_{{\bf b}\equiv (b_1,...,b_l)\in {\cal B}_{\cal A}(n)}\left({\bf q}_{\cal A}(n,{\bf b})|\left(b_j\cdot {\bf q}_{\cal A}(n,{\bf b}_{[1,j-1]})\right)_{j=1,...,l}\right).
\end{equation}

For any binary string ${\bf b}\in {\cal B}_{\cal A}(n)$ with length $l$, recall that $I_{{\cal P},{\cal A}}(n,{\bf b})$ is the product of intersecting the entire space $I_{\cal P}(n)\equiv \mathbb Q^{D_{\cal P}(n)}$ with the $l$ half spaces $(b_j\cdot {\bf q}_{\cal A}(n,{\bf b}_{[1,j-1]}))$ for $j=1,...,l$. In the same sense that $I_{\cal P}(n)$ may be understood as being decomposable into the $I^S_{\cal P}(n,s)$'s, the ${\bf b}$-determined $I_{{\cal P},{\cal A}}(n,{\bf b})$ is in a sense decomposable into the $I^S_{{\cal P},{\cal A}}(n,s,{\bf b})$'s where each is the product of intersecting $I^S_{\cal P}(n,s)$ with the aforementioned $l$ half spaces. Of course, some of these $I^S_{{\cal P},{\cal A}}(n,s,{\bf b})$'s are already of dimensions below $D_{\cal P}(n)-1$ or even empty. However, the remaining fully-dimensional ones could still make up a substantial number. It is probably that this number would not get down that much from its initial $S_{\cal P}(n)$. Then, the algorithm ${\cal A}$ would face essentially as difficult a problem when confined to $I_{{\cal P},{\cal A}}(n,{\bf b})$ as when facing the entire $I_{\cal P}(n)$ in the beginning.

Also in view of the three-way categorization and the~(\ref{baba})-to-(\ref{mama}) solvability criterion, we have come to the following property which we believe is pivotal to an RILCOP being hard to tackle by any SIMPLE algorithm. Since the property is not about any particular ${\cal A}$, we need an update on the notations using binary strings. Given any nonzero $D_{\cal P}(n)$-dimensional sequence $({\bf q}_1,...,{\bf q}_l)$, let $I_{{\cal P},Q}(n,{\bf q}_1,...,{\bf q}_{l})$ be the product of intersecting $I_{\cal P}(n)$ with the $l$ half spaces $({\bf q}_j)$ for $j=1,...,l$. Let the $I^S_{{\cal P},Q}(n,s,{\bf q}_1,...,{\bf q}_l)$'s be similarly defined. 

\begin{definition}\label{def-resistant}
An RILCOP ${\cal P}$ would be considered {\em splinter-prone} when an exponential number of $I^S_{\cal P}(n,s)$'s are fully-dimensional; more importantly, as long as the number of fully-dimensional ones among the $I^S_{{\cal P},Q}(n,s,{\bf q}_1,...,{\bf q}_{p(n)})$'s is some exponential $e(n)$ after some polynomial $p(n)$ number of queries ${\bf q}_1,...,{\bf q}_{p(n)}$, using any ${\bf q}$ other than anything proportional to the ${\bf q}_j$'s that have already appeared in order to conduct another query on these $e(n)$ polyhedra would encounter some $e(n)-q(n)$ splinters where $q(n)$ is only polynomial. 
\end{definition}

A splinter-prone problem ${\cal P}$ would basically have most of its solution-induced polyhedral sets fall into category (spL) when being queried by a plane; it would leave only a polynomial of these sets to categories (sCp) and (Rsp). More precisely, even after undergoing intersections with a polynomial number of half spaces, all the exponential number of inequalities that help define the problem's exponential number of solution sets would only help lead a polynomial number of these sets to reside decidedly within one side of a given query plane, in either the (sCp) or (Rsp) form. The vast majority would appear on both sides in the (spL) form.  

\begin{theorem}\label{doable}
A splinter-prone RILCOP ${\cal P}$ would not be polynomial-time solvable by any ${\cal P}$-compatible SIMPLE algorithm ${\cal A}$. 
\end{theorem}

\noindent{\bf Proof of Theorem~\ref{doable}: }Before any query, ${\cal P}$ certainly has an exponential $e_0(n)\leq S_{\cal P}(n)$ number of fully-dimensional solution sets that have not been fully scraped. By Definition~\ref{def-solved}, the problem has not been solved.

Suppose after some polynomial $p(n)$ number of queries, there still exist some exponential $e_{p(n)}(n)\equiv e_0(n)-q_1(n)-\cdots-q_{p(n)}(n)$ number of fully-dimensional sets resulting from intersecting all ${\cal P}$'s $e_0(n)$ number of solution sets with some $p(n)$ half spaces which make up one out of $2^{p(n)}$ potential half-space streams induced by all queries in the past, where $q_1(\cdot),...,q_{p(n)}(\cdot)$ are all polynomial; also, none of the sets has been fully scraped. Note $q(\cdot)\equiv q_1(\cdot)+\cdots+q_{p(\cdot)}(\cdot)$ would still be polynomial.
 
Consider an arbitrary $(p(n)+1)$-th query that is not proportional to  any of the past $p(n)$ queries. Since ${\cal P}$ is splinter-prone in the sense of Definition~\ref{def-resistant}, there would be some polynomial $q^0(n)$ number of the $e_{p(n)}(n)$ sets that have the (sCp) or (Rsp) relations with the query plane; each of the remaining $e_{p(n)}(n)-q^0(n)$ sets is splintered into two. If we choose either side of the plane, some polynomial $q_{p(n)+1}(n)$ number of the (sCp)- and (Rsp)-sets would be on one side and the other $q^0(n)-q_{p(n)+1}(n)$ of them on another, where $q_{p(n)+1}(n)\leq q^0(n)$; meanwhile, either side would have $e_{p(n)}-q^0(n)$ sets resulting from (spL). On the side with $q^0(n)-q_{p(n)+1}(n)$ (sCp)- and (Rsp)-sets, the total number of post-query sets would be  $e_{p(n)}-q^0(n)+q^0(n)-q_{p(n)+1}(n)=e_{p(n)}(n)-q_{p(n)+1}(n)$.  

That is, an exponential $e_{p(n)+1}(n)\equiv e_{p(n)}-q_{p(n)+1}(n)\equiv e_0(n)-q_1(n)-\cdots-q_{p(n)}(n)-q_{p(n)+1}(n)$ number of fully-dimensional sets would be deducible from intersecting ${\cal P}$'s solution sets with $p(n)+1$ half spaces which make up one out of $2^{p(n)+1}$ potential half-space streams induced by all $p(n)+1$ past queries; also, none of the sets would have been fully scraped. Note $q'(\cdot)\equiv q(\cdot)+q_{p(\cdot)+1}(\cdot)\equiv q_1(\cdot)+\cdots+q_{p(\cdot)+1}(\cdot)+q_{p(\cdot)+1}(\cdot)$ would remain polynomial.  

From the above induction, we see that after any polynomial $p(n)$ number of queries, there would remain some exponential $e_0(n)-q(n)$ number of fully-dimensional and yet not fully scraped sets for some polynomial $q(\cdot)$ that result from intersecting ${\cal P}$'s original $e_0(n)$ solution sets to $p(n)$ half spaces which make up one out of $2^{p(n)}$ potential half-space streams induced by the $p(n)$ queries. By Definition~\ref{def-solved}, ${\cal P}$ has not been solved after these $p(n)$ queries. 

Note faces aligned with any of the $p(n)$ past queries have been scraped already, there would not be any additional scraping by letting  the $(p(n)+1)$-th query to repeat a past one. 
By the arbitrariness of $p(\cdot)$, the above would mean that ${\cal P}$ is not polynomial-time solvable. \qed 

In view of Theorem~\ref{doable}, the polynomial-time solvable AP must not be splinter-prone. 



\section{The Traveling Salesman Problem}\label{sec5}

For TSP, recall that $D_{\mbox{tsp}}(n)=n\cdot (n+1)$ and $S_{\mbox{tsp}}(n)=n!$ for each $n\in \mathbb N$. We can name the $n!$ permutations of $\{1,2,...,n\}$ as the $\bar{\pi}(n,s)$'s for $s=0,1,...,n!-1$ where each $\bar{\pi}(n,s)\equiv (\bar{\pi}(n,s,m))_{m=1,2,...,n}$; also, we can let $\bar{\pi}(n,0)$ be the identity permutation with each $\bar{\pi}(n,0,m)=m$; in addition, it is possible to let $\bar{\pi}(n,s)$ be lexicographically before $\bar{\pi}(n,s+1)$ for $s=0,1,...,n!-2$ in the sense that $\bar{\pi}(n,s,m)\geq \bar{\pi}(n,s+1,m)+1$ for some $m=2,...,n$ only when $\bar{\pi}(n,s,m')\leq \bar{\pi}(n,s+1,m')-1$ for some $m'=1,...,m-1$. With each vector ${\bf y}$ in $\mathbb Q^{n\cdot (n+1)}$ understood as $(y_{01},y_{02},...,y_{0n},y_{10},y_{12},...,y_{1n},...,y_{n0},y_{n1},...,y_{n,n-1})^T$, we can let each ${\bf x}_{\mbox{tsp}}(n,s)$ be the one whose $(0,\bar\pi(n,s,1))$, $(\bar{\pi}(n,s,1),\bar\pi(n,s,2))$, $...$, $(\bar{\pi}(n,s,n-1),\bar{\pi}(n,s,n))$, and $(\bar{\pi}(n,s,n),0)$ components are ones and other components are zeros. 

\begin{proposition}\label{p-easy}
For TSP, every solution set $I^S_{\mbox{tsp}}(n,s)$ is fully-dimensional and every query ${\bf x}_{\mbox{tsp}}(n,s)-{\bf x}_{\mbox{tsp}}(n,s')$ generates a face for $I^S_{\mbox{tsp}}(n,s)$.
\end{proposition}
\noindent{\bf Proof of Proposition~\ref{p-easy}: }Due to symmetry, let us just suppose $s=0$. For the first claim, consider the instance ${\bf c}\equiv (c_{01},c_{02},...,c_{0n},c_{10},c_{12},...,c_{1n},...,c_{n0},c_{n1},...,c_{n,n-1})^T\in \mathbb Q^{n\cdot(n+1)}$ where $c_{01}=c_{12}=c_{23}=\cdots=c_{n-1,n}=c_{n0}=0$ and every other $c_{ij}=1$. It clearly satisfies
\[\left({\bf x}_{\mbox{tsp}}(n,0)-{\bf x}_{\mbox{tsp}}(n,s)\right)^T\cdot{\bf c}<0,\hspace*{.8in}\forall s=1,2,...,n!-1.\] 

For the second claim, consider any arbitrary $s=1,2,...,n!-1$. Now let ${\bf c}\in \mathbb Q^{n\cdot (n+1)}$ be such that $c_{0,\bar{\pi}(n,s',1)}=c_{\bar\pi(n,s',1),\bar{\pi}(n,s',2)}=\cdots=c_{\bar{\pi}(n,s',n-1),\bar{\pi}(n,s',n)}=c_{\bar{\pi}(n,s',n),0}=0$ for both $s'=0$ and $s'=s$ while any other $c_{ij}=1$. Clearly, 
\[\begin{array}{l}
\left({\bf x}_{\mbox{tsp}}(n,0)-{\bf x}_{\mbox{tsp}}(n,s)\right)^T\cdot {\bf c}=0\;\mbox{ and yet}\\
\;\;\;\;\;\;\;\;\;\mbox{ for any }s'\mbox{ other than }0\mbox{ or the chosen }s,\;\left({\bf x}_{\mbox{tsp}}(n,0)-{\bf x}_{\mbox{tsp}}(n,s')\right)^T\cdot{\bf c}<0.
\end{array}\] 
As long as the equation remains, the above inequality could be maintained even after each $c_{ij}$ has been perturbed by up to $1/(2n+2)$. \qed

All $I^S_{\mbox{tsp}}(n,s)$'s are also balanced. Indeed, more can be said. The $n\cdot(n+1)$ components of every ${\bf x}_{\mbox{tsp}}(n,s)-{\bf x}_{\mbox{tsp}}(n,s')$ would contain an equal number of $+1$'s and $-1$'s, while the rest are 0's. Table~\ref{tab1} provides an illustration. 

\begin{table}
\begin{center}
\caption{An Illustration of Queries that Could Form a Polyhedron}
\begin{tabular}{||c|c|c|c|c|c|c||}\hline\hline
$(i,j)$ & ${\bf x}_{\mbox{tsp}}(n,0)$ & $\ldots\ldots$ & ${\bf x}_{\mbox{tsp}}(n,0)$  & ${\bf q}_1$ & $\cdots$ & ${\bf q}_{p(n)}$ \\ 
 & $\;\;\;\;\;\;-{\bf x}_{\mbox{tsp}}(n,1)$ & & $\;\;\;\;\;\;-{\bf x}_{\mbox{tsp}}(n,n!-1)$ & & & \\ \hline
$(0,1)$ & $+1-1=0$ & $\cdots\cdots$ & $+1$ & $q_{1,01}$ & $\cdots$ & $q_{p(n),01}$ \\
$(0,2)$ & 0 &  $\cdots\cdots$  & 0 & $q_{1,02}$ & $\cdots$ & $q_{p(n),02}$ \\
$\cdots$ & $\cdots$ &   $\cdots\cdots$  & $\cdots$ & $\cdots$ & $\cdots$ & $\cdots$ \\
$(0,n)$ & 0 & $\cdots\cdots$   & $-1$ & $q_{1,0n}$ & $\cdots$ & $q_{p(n),0n}$ \\
$(1,0)$ & 0 &  $\cdots\cdots$  & $-1$ & $q_{1,10}$ & $\cdots$ & $q_{p(n),10}$ \\
$(1,2)$ & $+1-1=0$ &  $\cdots\cdots$  & $+1$ & $q_{1,12}$ & $\cdots$ & $q_{p(n),12}$ \\
$\cdots$ & $\cdots$ &  $\cdots\cdots$  & $\cdots$ & $\cdots$ & $\cdots$  & $\cdots$ \\
$(1,n)$ & 0 &  $\cdots\cdots$  & 0 & $q_{1,1n}$ & $\cdots$ & $q_{p(n),1n}$ \\
$\cdots$ & $\cdots$ &  $\cdots\cdots$  & $\cdots$ & $\cdots$ & $\cdots$ & $\cdots$ \\
$(n,0)$ & $+1$ &  $\cdots\cdots$  & $+1$ & $q_{1,n0}$ & $\cdots$ & $q_{p(n),n0}$ \\
$(n,1)$ & 0 & $\cdots\cdots$  & 0 & $q_{1,n1}$ & $\cdots$ & $q_{p(n),n1}$ \\
$\cdots$ & $\cdots$ &  $\cdots\cdots$  & $\cdots$ & $\cdots$ & $\cdots$ & $\cdots$ \\
$(n,n-1)$ & $-1$ &  $\cdots\cdots$  & $-1$ & $q_{1,n,n-1}$ & $\cdots$ & $q_{p(n),n,n-1}$ \\ \hline\hline
\end{tabular}\label{tab1}\end{center}\end{table}

The same is true for AP. In view of Proposition~\ref{p-balanced}, it would probably not come as a surprise that the Hungarian method proposes only balanced queries. While the objective of AP is $\sum_{i=1}^n c_{i\pi(i)}$, to conform to it that of TSP can be written as $\sum_{i=0}^n c_{0,\pi((\pi^{-1}(n)+1)\mbox{ mod }(n+1)}$ where $\pi(0)$ is treated as 0. One reason that AP is much easier to solve than TSP might be that unlike the $({\bf x}_{\mbox{ap}}(n,0)-{\bf x}_{\mbox{ap}}(n,s))$'s which span a large space of balanced vectors, the $({\bf x}_{\mbox{tsp}}(n,0)-{\bf x}_{\mbox{tsp}}(n,s))$'s would probably leave many corners uncovered. Let us couch our more precisely worded and potentially more operable conjecture on Theorem~\ref{doable}. 

\begin{conjecture}\label{keystone}
TSP is splinter-prone in the sense of Definition~\ref{def-resistant}. 
\end{conjecture}

\noindent{\bf Incomplete Proof of Conjecture~\ref{keystone}: }First, it is clear by Proposition~\ref{p-easy} that every $I^S_{\mbox{tsp}}(n,s)$ is fully-dimensional. So we can safely let $e_0(n)$ be the exponential $n!$.

The next attempt should be on showing that for any polynomial $p(\cdot)$, there would exist another polynomial $q(\cdot)$ so that for any $n\in\mathbb N$, any queries ${\bf q}_1,...,{\bf q}_{p(n)}\in\mathbb Q^{n\cdot (n+1)}$, and any other query ${\bf q}\in\mathbb Q^{n\cdot(n+1)}$, the following would be true:\\
\indent\M almost all but at most $q(n)$ of the $s$'s that are with some ${\bf c}(s)\in \mathbb Q^{n\cdot(n+1)}$ to satisfy
\[ \left({\bf x}_{\mbox{tsp}}(n,s)-{\bf x}_{\mbox{tsp}}(n,s')\right)^T\cdot {\bf c}(s)<0,\hspace*{.8in}\forall s'=0,...,s-1,s+1,...,n!-1,\]
and ${\bf q}_m^{\;T}\cdot {\bf c}(s)<0$ for any $m=1,...,p(n)$, would contribute some ${\bf c}_{+1}(s),{\bf c}_{-1}(s)\in \mathbb Q^{n\cdot (n+1)}$ such that both satisfy ${\bf c}(s)$'s inequalities though not necessarily strictly; however, 
\[ {\bf q}^T\cdot {\bf c}_{+1}(s)<0,\;\;\;\;\;\;\mbox{ and yet }\;\;\;\;\;\;{\bf q}^T\cdot {\bf c}_{-1}(s)>0.\]

Right now we have no further clues on tackling this huge task! \qed



If Conjecture~\ref{keystone} were true, we would derive from Theorem~\ref{doable} that TSP is not polynomial-time solvable in the sense of Definition~\ref{def-solved}. Since TSP is an $NP$-equivalent problem \cite{Garey and Johnson}\cite{Karp}, one can prove $P\neq NP$ by showing that no polynomial-time algorithm can solve TSP. Because TSP belongs to those RILCOPs to which SIMPLE algorithms are of extreme importance, proving even this weaker property 
would already seem to have done a great deal to the ultimate proof of the $P-NP$ conjecture.

\section{Concluding Remarks}\label{sec6}

We have provided our understanding on how certain RILCOPs are solved by SIMPLE algorithms. Out of this arises the splinter-proneness notion that is intimately linked to the time complexity question. The conjecture that TSP is splinter-prone is so far not proved yet. Since it is an $NP$-equivalent problem or one of those easiest $NP$-hard problems, the potential conclusion could have quite far-reaching implications on the $P\neq
NP$ question. 


\end{document}